\documentclass{emulateapj} 
\usepackage{psfig}
\usepackage{apjfonts}
\usepackage{mathrsfs}
\shorttitle{FORS1 observations of the double MS of $\omega$ Cen}
\shortauthors{Sollima et al.}

\begin{document}

\title{Deep FORS1 observations of the double Main Sequence of $\omega$
Centauri\footnote{Based on FORS1 observations collected with the Very Large 
Telescope at the European Southern Observatory, Cerro Paranal, Chile, 
within the observing program 74.D-0369(B).}}

\author{A. Sollima\altaffilmark{1}, F. R.
Ferraro\altaffilmark{1}, M. Bellazzini\altaffilmark{2}, L.
Origlia\altaffilmark{2}, O. Straniero\altaffilmark{3} and E. Pancino\altaffilmark{2}}

\altaffiltext{1}{Dipartimento di Astronomia, Universit\`a di Bologna, via Ranzani 1, Bologna, Italy}
\altaffiltext{2}{INAF Osservatorio Astronomico di Bologna, via Ranzani 1, I-40127 Bologna, Italy}
\altaffiltext{3}{INAF Osservatorio Astronomico di Collurania, via M. Maggioni, I-64100 Teramo, Italy}

\begin{abstract}
We present the results of a deep
photometric survey performed with FORS1@VLT aimed at investigating the complex
Main Sequence structure of the stellar system $\omega$ Centauri. We confirm the
presence of a double Main Sequence and identify its blue component (bMS) over
a large field of view up to 26' from the cluster center. 
We found that bMS stars are significantly more concentrated toward the cluster center 
than the other "normal" MS stars. The bMS morphology and its position in the CMD
have been used to constrain the helium overabundance required to explain the observed MS morphology.
\end{abstract}


\section{Introduction}

The understanding of the origin and evolution of the stellar system $\omega$
Centauri (NGC5139) still represents one of the most intriguing unanswered
questions of stellar astrophysics. Unique among Galactic
star clusters in terms of structure, kinematics and stellar content, 
it is the only known globular cluster (GC) which shows a clear metallicity 
spread (Norris et al. 1996 and references therein).
Recent photometric surveys have revealed the presence of multiple sequences in
its color-magnitude diagram (CMD). In
particular, high-precision photometric analyses have revealed a discrete 
structure of its red giant branch
(RGB, Rey et al. 2004; Sollima et al. 2005a), indicating a complex star
formation history. Beside the dominant metal-poor
population (MP, ${\rm [Fe/H]} \sim -1.6$), three
metal-intermediate (MInt) components (spanning a range
of metallicity  $-1.3 < {\rm [Fe/H]} < -1.0$) and an
extreme metal-rich population (${\rm [Fe/H]} \sim -0.6$, Pancino et al. 2002) have
been identified. The different RGB populations of $\omega$~Cen
share different structural and dynamical properties 
(Norris et al. 1997; Ferraro et al. 2002; Pancino et al. 2003; Sollima et al. 2005a). In particular, Norris et al. (1997)
found that the 20 \% metal-rich tail of the distribution 
is more concentrated toward the cluster
center than the dominant metal-poor component.

Finally, Anderson (2002) and Bedin et al. (2004) discovered new peculiarities also along the Main
Sequence (MS) of 
the cluster. Indeed,
an additional blue MS (bMS, comprising $\sim$ 30\% of
the whole cluster MS stars) running parallel to the dominant one, has been
resolved. According to stellar models with canonical chemical abundances, the location
of the observed bMS would suggest a very low metallicity ($[Fe/H] < -2$). Conversely, the 
spectroscopic analysis of a sample of MS stars belonging
to the two MS components showed that bMS stars present a
metallicity $\sim 0.3~dex$ higher than that of the dominant cluster population (Piotto et al. 2005).
Norris (2004) suggested that a large helium overabundance ($\Delta Y \sim 0.15$) could
explain the anomalous position of the bMS in the CMD. However, such a large
helium abundance spread poses 
serious problems in the overall interpretation of the chemical enrichment history of this 
stellar system.

In this paper we present deep BR photometry\footnote{The entire catalog is only available in 
electronic form at the CDS via http://cdsweb.u-strasbg.fr/} covering a wide area 
extending from 6' to 26' from the cluster center with the aim of studying the
morphology, the radial extent and the distribution of the bMS population. 

\section{Observations and data reduction}
\label{redums}

\begin{figure}
\plotone{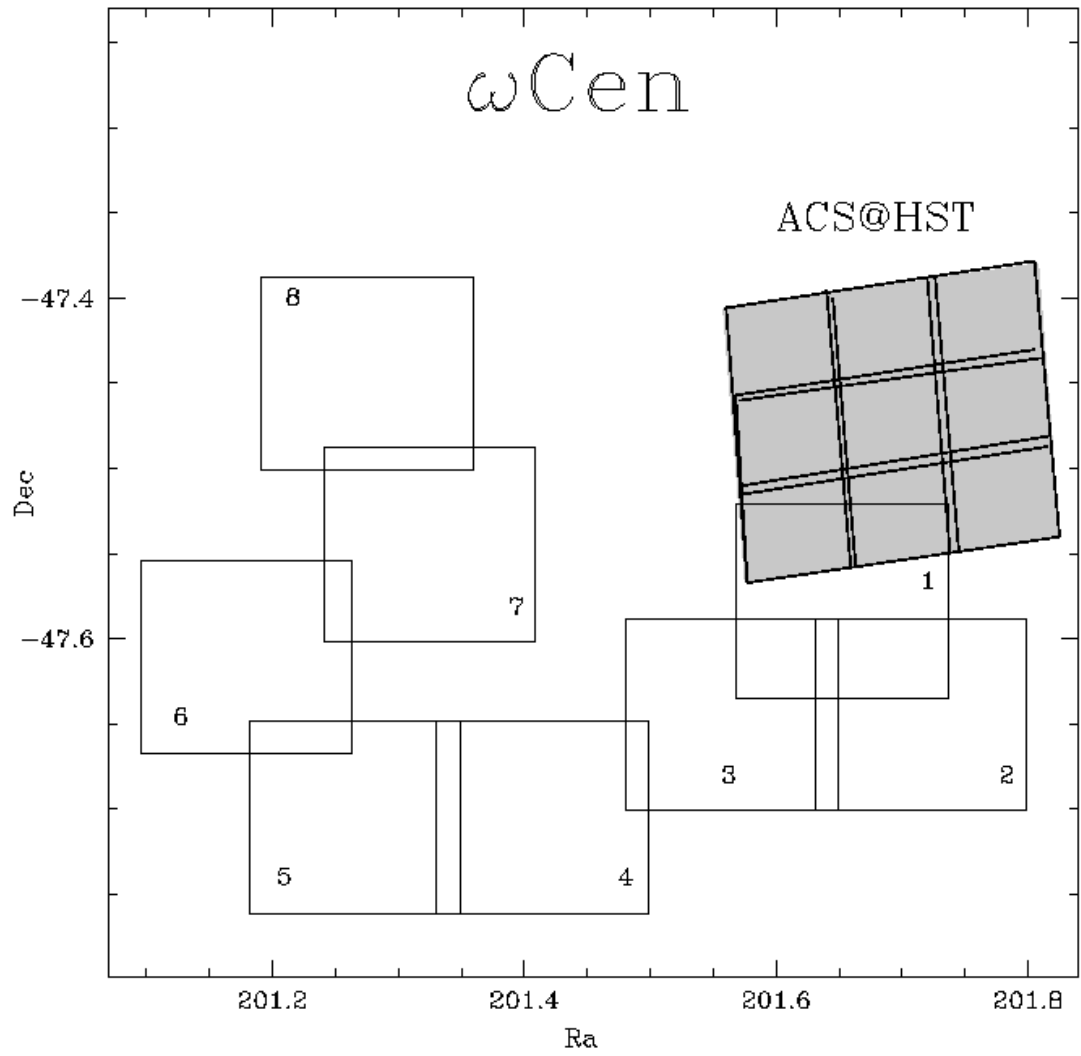}
\caption{Map of the region sampled by the FORS1 observations. North is up, east on
the right. The eight fields observed with FORS1 are shown. Grey boxes indicate the position of
the ACS photometry by Ferraro et al. (2004). The cluster center is located in the
central ACS field.} 
\label{mapms}
\end{figure}

The photometric data were obtained with the FORS1 camera, mounted at the Unit1
(UT1) of the ESO Very Large Telescope (VLT, Cerro Paranal, Chile). 
Observations were performed during 6 nights on March and April 2005 (see Table
\ref{obs_logms}), using the standard resolution mode of FORS1. In this
configuration the image scale is 0.2" $pixel^{-1}$ and the camera has a
global field of view of $6.8' \times 6.8'$ . A mosaic of 8 partially overlapping fields 
 spanning a wide area from 6' to 26' from the cluster center were observed (see Fig.
\ref{mapms}). The innermost field partially overlaps the deep ACS photometry described in
Ferraro et al. (2004) allowing linkage between the two datasets. 
The standard pre-reduction procedure was followed to remove the bias and to
apply flat-field corrections. We used 
the point-spread-function 
(PSF) fitting package DoPhot (Schechter et al. 1993) to obtain instrumental magnitudes for 
all the stars detected in each frame. For each field, two different frames were
observed through the B and R filters. The photometric analysis was performed 
independently on each image. Only stars detected in all the four frames were included in the final catalog. 
For each passband, the obtained magnitudes were transformed to
the same instrumental scale and averaged. As usual, the most isolated and
brightest stars in the field were used to link the aperture
magnitudes to the fitting instrumental ones, after normalizing for exposure time
and correcting for airmass. During the observing run, nine standard stars from the
Landolt (1992) list were observed. Aperture photometry was performed on each standard, and
used to derive the equations linking the aperture photometry to the standard
photometric system.
The calibration equations linking the b and r instrumental magnitudes to the standard system
ones (B and R) are:
$$ B-b=-0.067~(b-r)+27.123 $$
$$ R-r=0.007~(b-r)+27.400 $$
Both the slope and the zero points of the above relations are in good agreement with
those provided by the FORS1 support team.
Finally, a catalog with more than 70,000 calibrated stars was produced.

\section{Color Magnitude Diagram}

\begin{figure}
\plotone{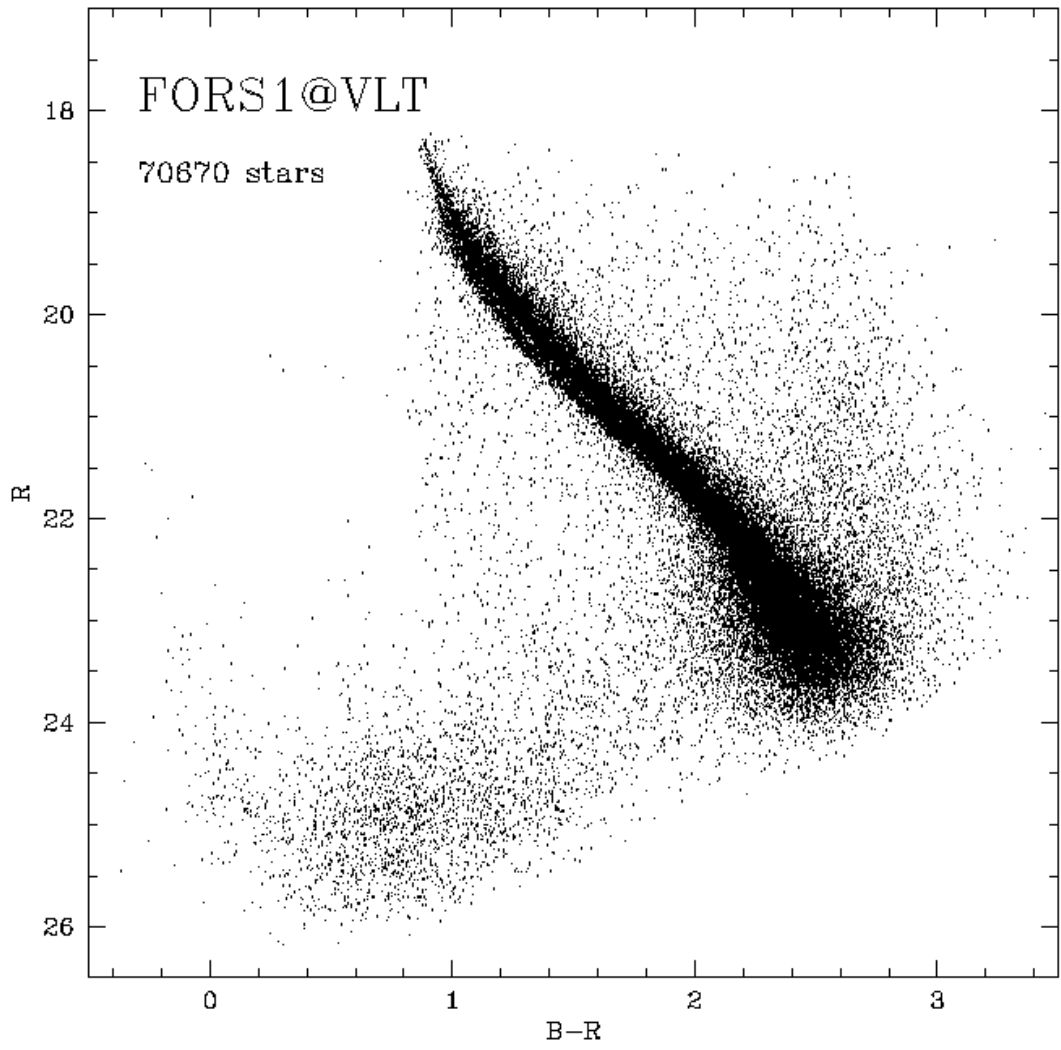}
\caption{R, B-R CMD for the global sample of
$\sim$70,000 stars measured in $\omega$ Cen.} 
\label{cmdms}
\end{figure}

Fig. \ref{cmdms} shows the (R, B-R) CMD of the entire sample. As can be seen, the
unevolved population of $\omega$ Cen down to $R\sim24$ is sampled. In
particular, two different MS populations can be identified:
\begin{itemize}

\item{The dominant MS population (rMS) containing $\sim 75\%$ of the entire
MS population stars;}
\item{A narrow blue MS, running parallel to the rMS population, can be
distinguished at $19.5<R<21$.}

\end{itemize}

\begin{figure}
\plotone{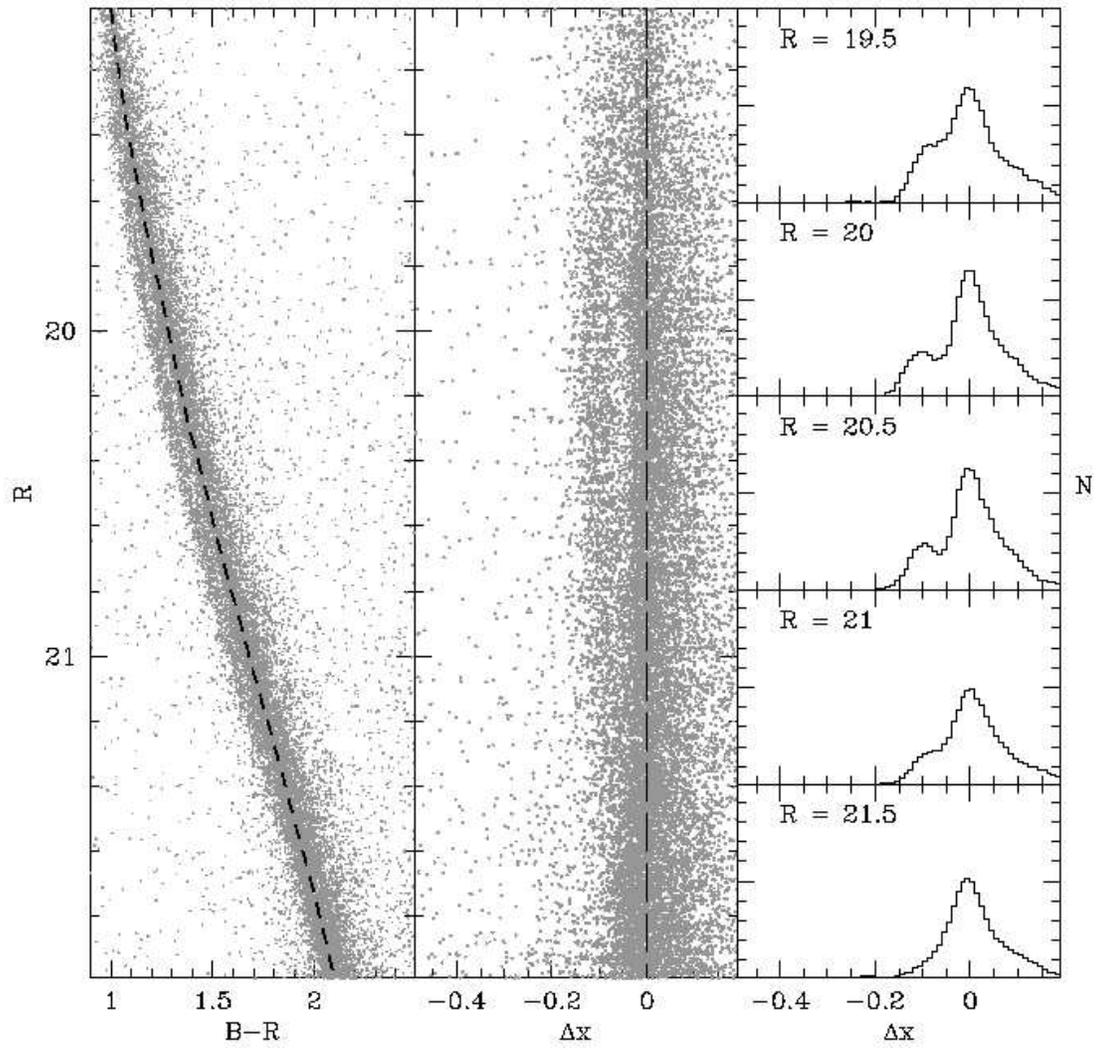}
\caption{The zoomed (R, B-R) CMD of $\omega$ Cen 
in the bMS region is shown in the $left~panel$. The mean ridge line of the rMS
population is overplotted to the CMD as a dashed line. The distribution of  
MS stars with respect to the reference mean ridge line is shown in the $middle~panel$. In the 
$right~panel$ the histograms of the distances from the mean ridge line at
different magnitude levels are shown.} 
\label{msdist}
\end{figure}

\begin{figure}
\plotone{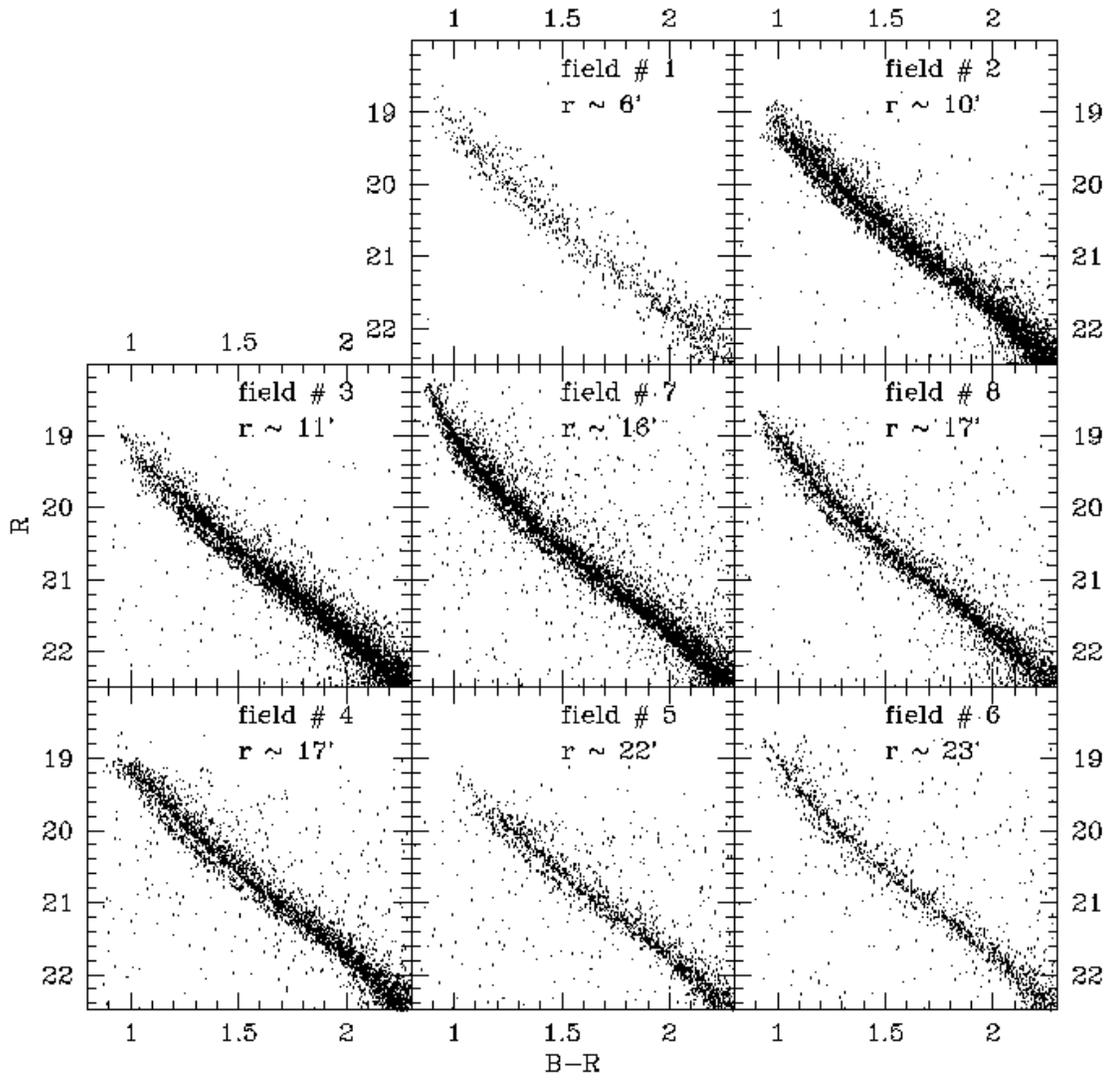}
\caption{The (R, B-R) CMDs of $\omega$ Cen zoomed 
in the bMS region for each of the eight observed fields. The field numbers are the same of Fig.
\ref{mapms} .} 
\label{allms}
\end{figure}

In order to investigate the morphology and properties of the bMS, we studied the
distance distribution of bMS stars from the mean 
ridge line of the rMS. 
The rMS mean ridge line was computed by 
selecting by eye the stars belonging to the dominant MS component. We took care in
excluding both blue objects populating the faint part of the CMD (i.e. white
dwarfs) and the bright red population (mostly due to the Galactic field). 
Particular care was taken in excluding bMS stars which could contaminate
the selected rMS sample. However, bMS stars are well separated from the rMS over
a large magnitude range.
We estimate the contamination from bMS stars to be $< 10$\% between
$19<R<21$, with negligible impact on the ridge line determination. 
Then, we fitted the selected stars with a low-order polynomial and rejected stars lying at a 
$>~2 \sigma$ distance. The procedure was iterated until convergence to a stable fit was obtained.
Then, we defined the observable $\Delta x$ as the geometrical distance of each MS
star from the reference ridge line. Fig. \ref{msdist} shows the obtained mean ridge line and the
distance distributions in the $left$ and $middle~panels$,
respectively. In the $right~panel$ the histograms of the distances from the mean ridge line at
different magnitude levels are shown. As can be noted, the bMS appears distinguishable from the rMS at
magnitude $R\sim 19.2$, reaching a maximum separation from the rMS at $R\sim20.3$ . 
The bMS merges into the bulk of the cluster MS at a fainter magnitude ($R\sim21.3$).
Fig. \ref{allms} shows the CMD for each of the eight fields observed in the present analysis.
The CMDs related to the innermost fields are significantly less populated
than the outer ones. This effect is due to the presence of many bright saturated stars that
cover most of the chip field of view thus allowing a meaningful photometric
analysis only over a small fraction of the chip area.  
As can be seen, the MS splitting is visible in all the CMDs regardless of the distance from the
cluster center. This evidence indicates that bMS stars cover the entire extension of the
cluster, being part of the population mix of $\omega$ Cen. 
The evident MS splitting clearly visible in Fig. \ref{cmdms}, \ref{msdist} and \ref{allms} confirms what
already found by Anderson (2002) and Bedin et al. (2004) on the basis of high-precision HST
photometric studies performed on two small fields located in a peripheral region of the cluster (at $\sim$ 7' and 17' 
from the cluster center, respectively). However, this is the first time that the bMS 
is identified over such a large area. This allowed us to perform a meaningful 
comparison between the bMS and rMS radial distributions.

\section{Radial Distributions}

\begin{figure}
\plotone{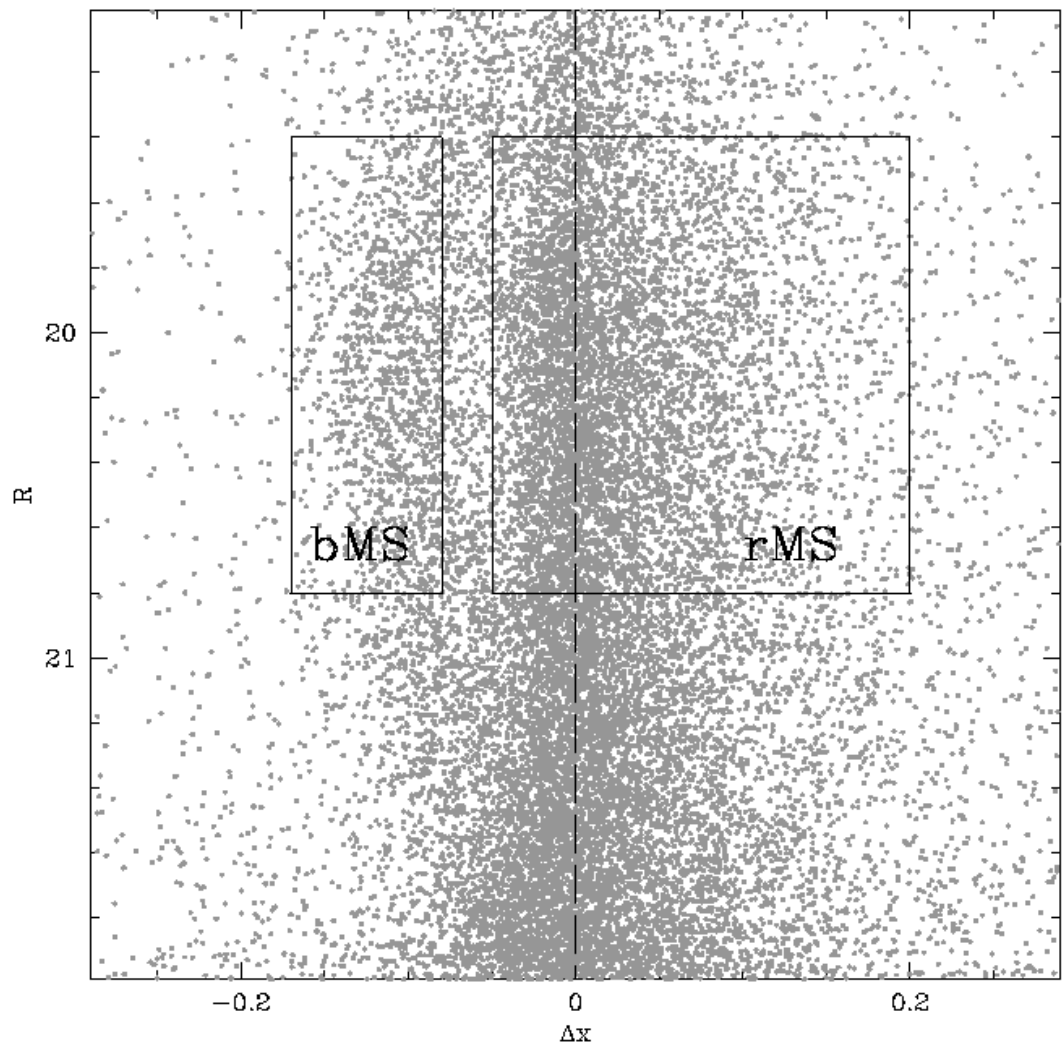}
\caption{Selection boxes for the rMS and bMS population in the ($\Delta x $,R) plane 
($Middle~panel$ of Fig. \ref{msdist}) .} 
\label{boxms}
\end{figure}

In order to derive the radial distributions of the two MS populations, 
we selected the samples of bMS and rMS stars on the basis of the distances from the 
mean ridge line of the rMS shown in the $middle~panel$ of Fig. \ref{msdist} . 
The adopted selection boxes for the two MS components are shown in Fig.~\ref{boxms} .
Only stars in the magnitude range $19.4<R<20.8$ have been used, in order
to limit the analysis to the region in which the two MS components are more clearly distinguished. 
For the rMS sample only stars with $-0.05 < \Delta x < 0.2$ have been considered, while bMS 
stars were selected among stars with $-0.17 < \Delta x < -0.08$. On the basis of
these selection criteria we isolated
7,122 rMS and 1,718 bMS stars. 
A residual contamination of rMS stars can be still present in the bMS selection
box\footnote{Of course the same effect produces an inverse contamination of bMS stars in the rMS selection box. We
neglected such an effect since its impact on the rMS sample is less than 0.5\% at any distance from the cluster 
center.}. This effect could be important expecially in the innermost fields, where severe 
crowding produces large uncertainties in magnitude and color. 
In order to quantify this effect, we performed an extensive set of experiments with artificial stars (Bellazzini
et al. 2002): a sample of 100,000 artificial stars
whose magnitude were extracted from the rMS mean ridge line have been 
simulated in the observed fields. The spread of the
distribution around the rMS
mean ridge line reflects the photometric errors in the different regions of the cluster.
Then, we counted the number of stars that satisfy the bMS selection criterion and evaluated
the contamination factor. The above analysis indicated that $\sim 7$\% of the bMS stars in the inner 12' are expected
to be spurious rMS stars. The contamination decreases rapidly at 
larger radii, becoming neglegible at $r>16'$. We took into account this effect in the 
following analysis. 

\begin{figure}
\plotone{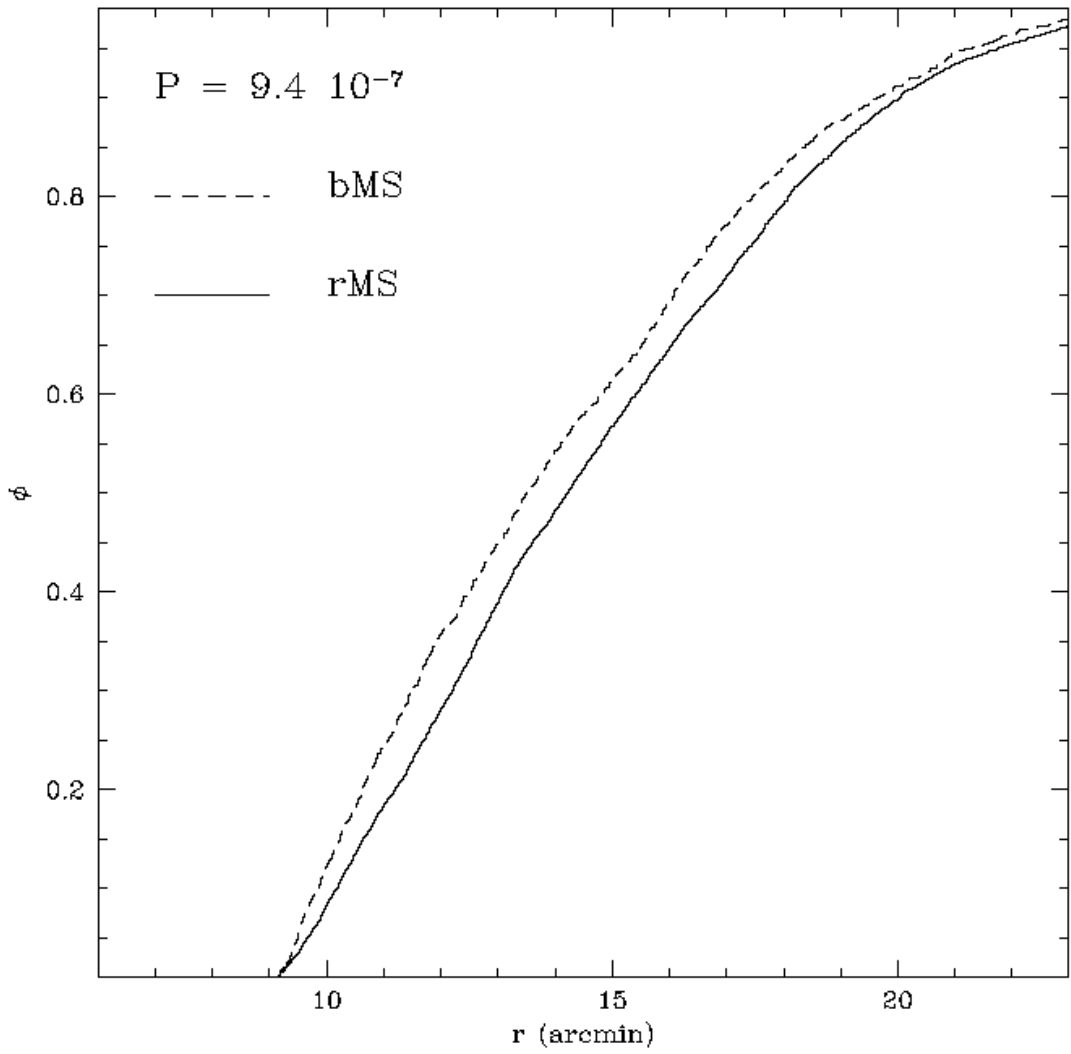}
\caption{Cumulative radial distribution for the rMS stars (solid line) and bMS stars (dashed
line) as a function of their projected distance from the cluster center.} 
\label{ksms}
\end{figure}

\begin{figure}
\plotone{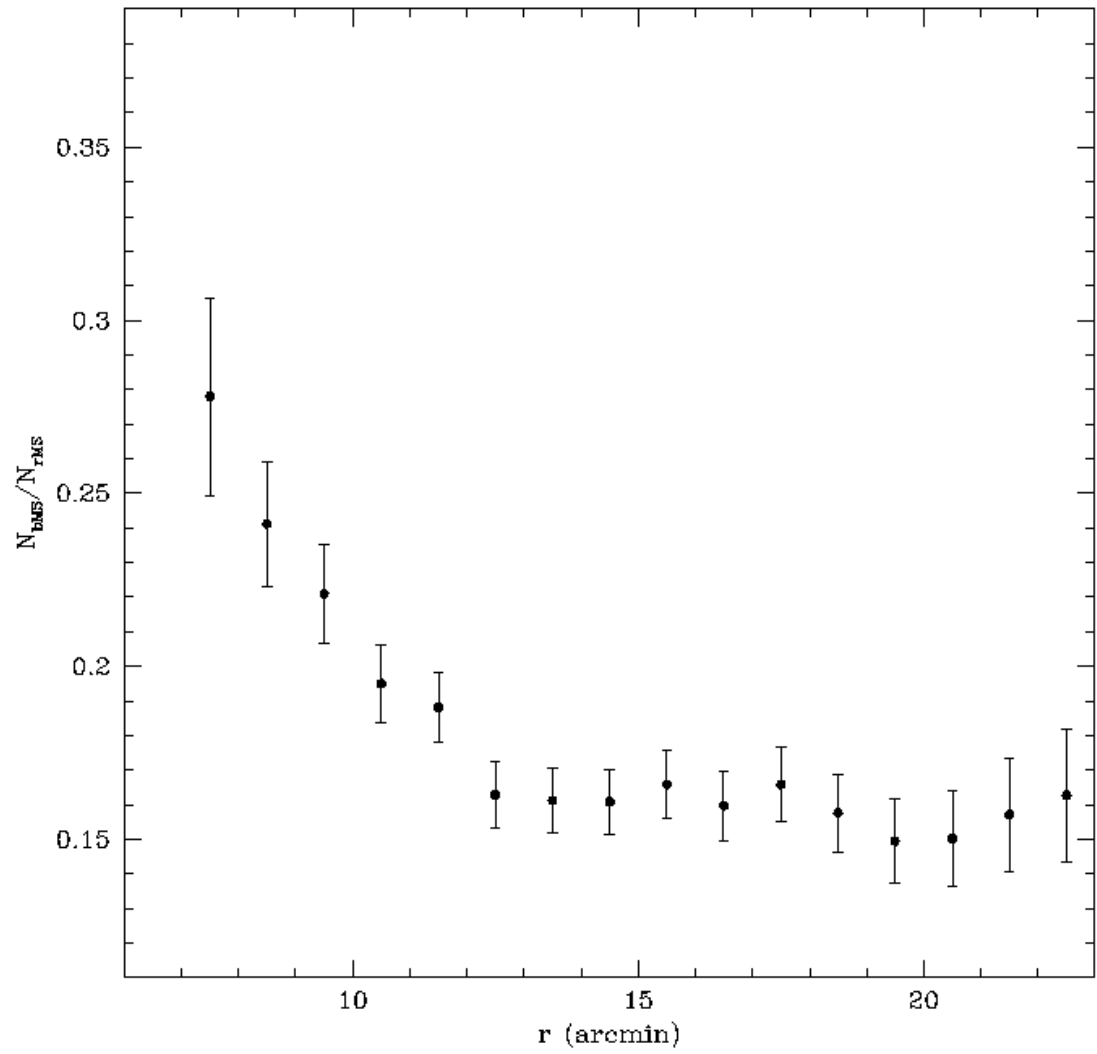}
\caption{Corrected ratio between the number of bMS and rMS 
stars at different distances from the cluster center.} 
\label{ratioms}
\end{figure}

\begin{figure}
\plotone{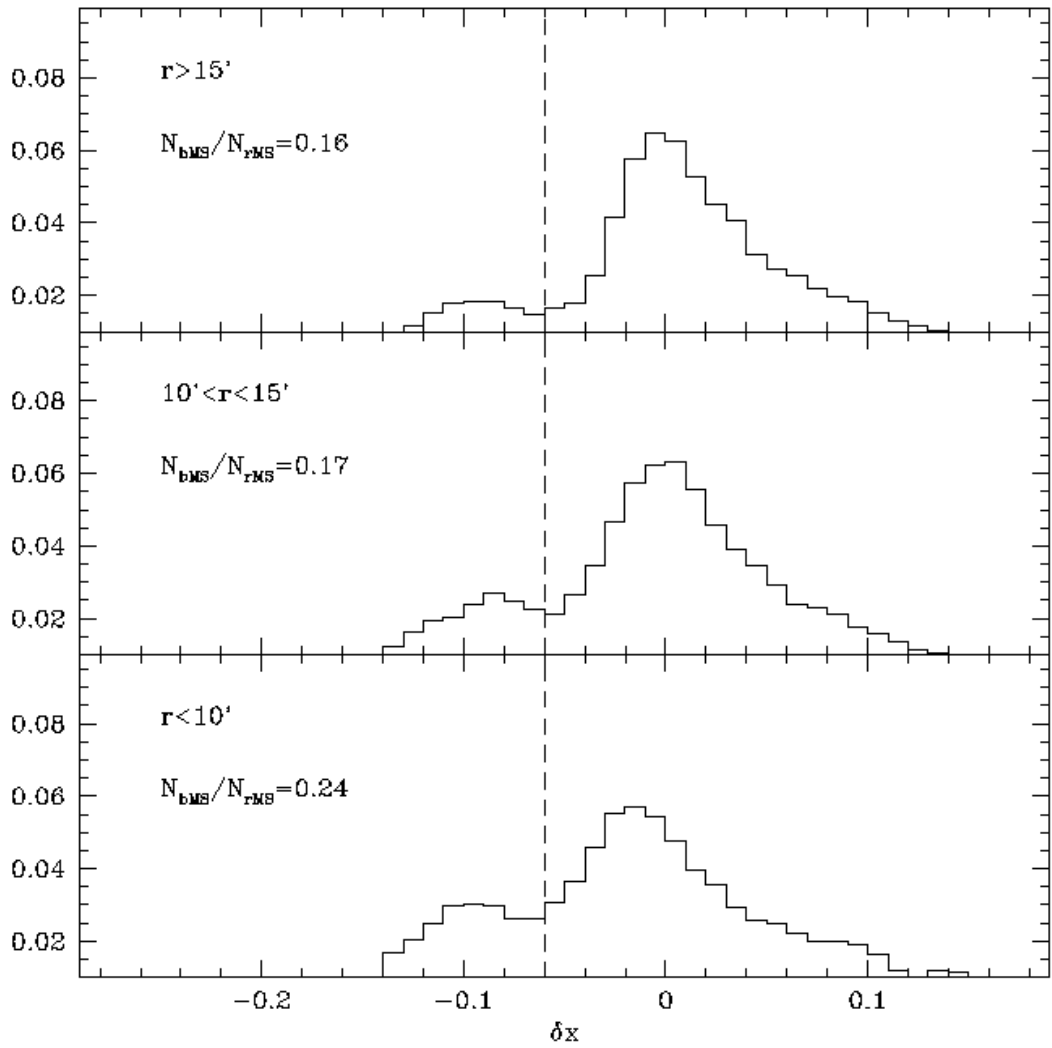}
\caption{Distribution of MS stars with respect to the reference mean ridge line calculated 
in the magnitude range $19.4<R<20.8$ for stars at distance $r<10'$ (bottom $panel$), 
$10'<r<15'$ (middle $panel$) and $r>15'$ (top $panel$) from the cluster center.} 
\label{dxdist}
\end{figure}

Fig. \ref{ksms} shows the cumulative radial distribution of the rMS and bMS
samples. The comparison indicates 
that the two MS components of $\omega$ Cen are distributed in a different way.
A two-dimensional generalization of the Kolmogorov-Smirnov test (Peacock 1983, 
Fasano \& Franceschini 1987) gives a probability that the spatial distribution 
of the bMS sample and the rMS one are drawn from the same parent distribution of less than 
$10^{-6}$. In particular, bMS stars are more concentrated toward the cluster center. 
To further investigate this effect, we computed the ratio between the number of bMS and rMS 
stars at different distances from the cluster center. This observable is insensitive to the
photometric completeness and it could provide important information on the relative
frequency of bMS stars. Fig. \ref{ratioms} shows the calculated ratio corrected for
contamination effects (see above) as a function of the distance from the
cluster center.  
On average, the bMS accounts for $\sim 24$\% of the whole MS population. 
As can be noted, the relative fraction of bMS-to-rMS stars decreases from 0.28
(at distances $\sim 7'$) to 0.15 (at distances $>19'$). 
This is best put into evidence in Fig. \ref{dxdist} where the 
distribution of MS stars with respect to the rMS mean ridge line, calculated 
in the magnitude range $19.4<R<20.8$ for stars at three different distances from the
cluster center, is shown.

\section{Comparison with Theoretical Isochrones}

Apart from the HB, the MS is the most helium-sensitive region of the CMD.
A significant helium overabundance could explain the anomalous location of the
bMS in the CMD because of the effect on the mean opacity that leads stars to
higher temperatures and bluer colors (Norris 2004).
The relative location of the two MS components shown in the CMD of Fig. 
\ref{cmdms} allowed us to infer an indirect estimate of the helium
content of the bMS population by means of a detailed comparison with suitable
theoretical models. In doing this, we used a new set of theoretical isochrones calculated adopting the most up-to-date
input physics and spanning a wide range in helium abundance. 
A detailed description of the evolutionary code used to compute these new models
can be found in Straniero et al. 1997.
To compare the observed CMD with theoretical isochrones, a
distance modulus and a reddening correction have to be adopted. In the following we used 
$(m-M)_0 = 13.70 \pm 0.13$ (Bellazzini et al. 2004). 
Concerning the reddening and extinction coefficients, we used $E(B-V)=0.11\pm
0.01$ (Lub 2002), $A_B=4.1~E(B-V)$, $A_R=2.32~E(B-V)$ (Savage \& Mathis 1979).

For the dominant rMS population we adopted a metallicity of $[Fe/H]=-1.6$ as 
suggested by the most extensive spectroscopic
surveys performed on giant stars (Norris et al. 1996; Suntzeff \& Kraft 1996), 
while for the bMS we adopted a significantly higher metallicity of $[Fe/H]=-1.3$,
as indicated by Piotto et al. (2005).
The contribution of
the $\alpha$-element enhancement has been taken into account by simply rescaling
standard models to the global metallicity [M/H], according to 
the following relation 
$$
 \mathrm{[M/H]}~=~\mathrm{[Fe/H]}+\log_{10}(0.638~10^{[\alpha/Fe]}+0.362)~~~~~\mbox{Salaris et al. (1993)}
$$

\begin{figure}
\plotone{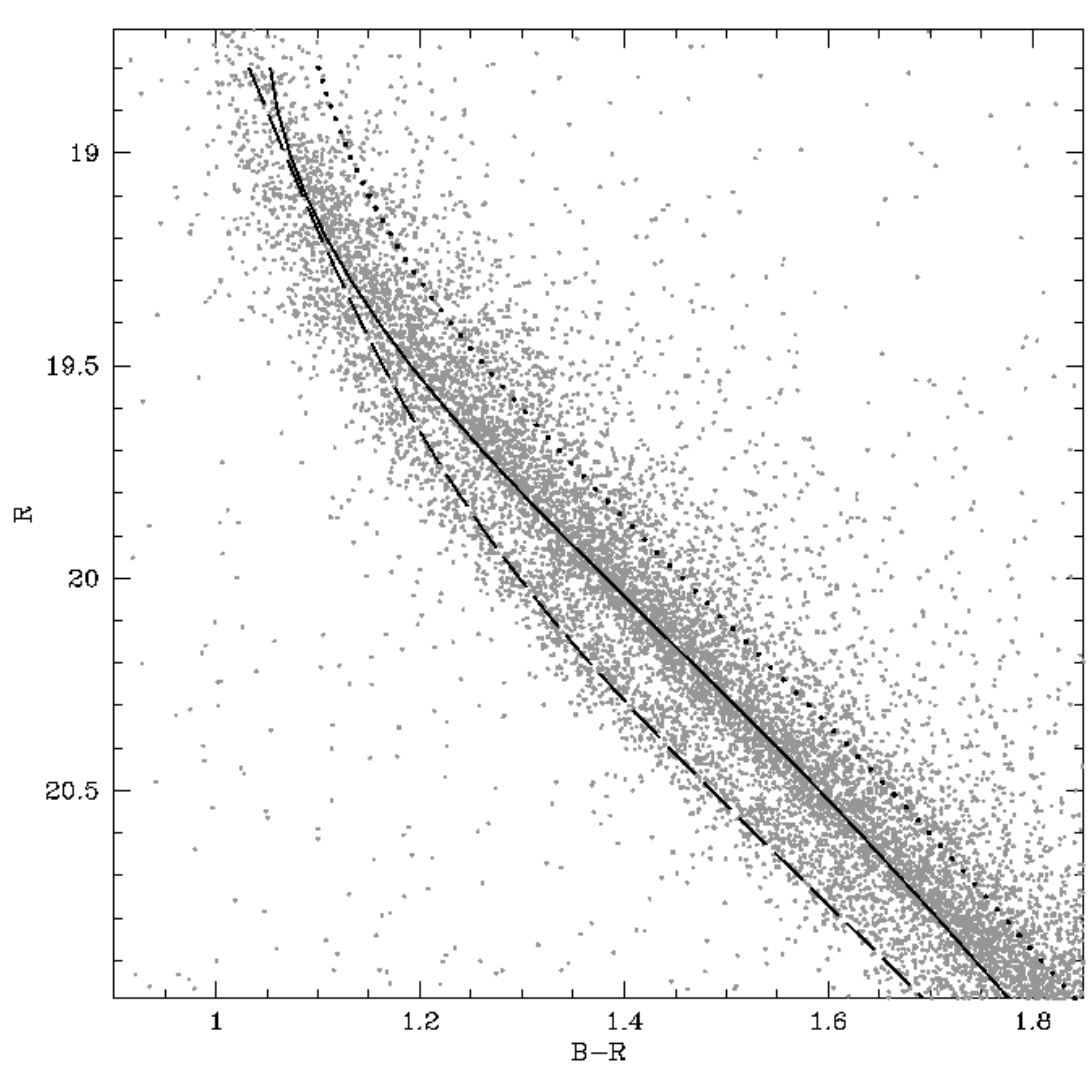}
\caption{Isochrone fitting of the MS populations of $\omega$ Cen.
Theoretical isochrones with appropriate metallicity and helium abundance
([Fe/H]=-1.6 ; Y=0.246: {\it solid line}, [Fe/H]=-1.3 ;  Y=0.246:{\it dotted
line} , Y=0.40: {\it dashed line}) are overplotted.} 
\label{msiso}
\end{figure}

We adopted $[\alpha/Fe] = +0.3$ for both the rMS and the
bMS samples, according to the most recent high-resolution
spectroscopic results (Norris \& Da Costa 1995; Smith et al. 2000; Vanture et
al. 2002). For the two MS components we computed a set of models with canonical helium abundance 
(Y=0.246, Salaris et al. 2004) and various helium enhancement levels.
The metallicity, in terms of mass fraction Z, has been
calculated according to the relation 
$$
Z = (1-Y)
\frac{10^{[M/H]+log[Z/X]_\odot}}{1+10^{[M/H]+log[Z/X]_\odot}}~~~~~~~~~~\mbox{Sollima et al. (2005b)}
$$
where $[Z/X]_\odot = 0.0176$, according to Lodders et al. (2003).
Fig. \ref{msiso} shows the isochrone fitting for the two observed MS
populations of $\omega$ Cen. As can be seen the metal-poor isochrone nicely fits 
the rMS. While the metal-rich isochrone with cosmological helium abundance
cannot reproduce the location of the bMS in the CMD, a significant helium enhancement 
($Y>0.35$) is required to reproduce the bMS. The bMS is best fitted by an
isochrone with Y=0.40 . This value is in good agreement with that predicted by
Norris (2004) and that estimated by Piotto et al. (2005) on the basis of ACS
observations of a peripheral region of the cluster. However, the exact amount of helium
overabundance needed to explain the observed MS morphology is still largely
uncertain because of the following reasons:

\begin{itemize}
\item{Uncertainties in the color-temperature conversion. A helium overabundance significantly alters the 
trasparency of the atmosphere. The adoption of color-temperature conversion based on standard models with no
helium overabundance produces a shift in color that mimics a larger helium
abundance;}
\item{Uncertainties on the metal abundances. Small changes in the iron and/or
$\alpha$-elements abundances produce significant effects on the MS morphology.
Theoretical models
indicate that a variation of $\Delta Z = 0.0001$ in the metal content difference between the two MS
populations would mimic the effect of a change of $\Delta Y=0.012$ in the deduced
helium abundance;}
\item{Theoretical uncertainties in the helium-rich isochrones. Until now, none of the known
stellar systems has been found to have a helium abundance $Y>0.30$ .
Therefore, there are no direct observational 
constraints that allow to confirm the effect of the large helium overabundance predicted by theoretical models.}
\end{itemize}

\begin{figure}
\plotone{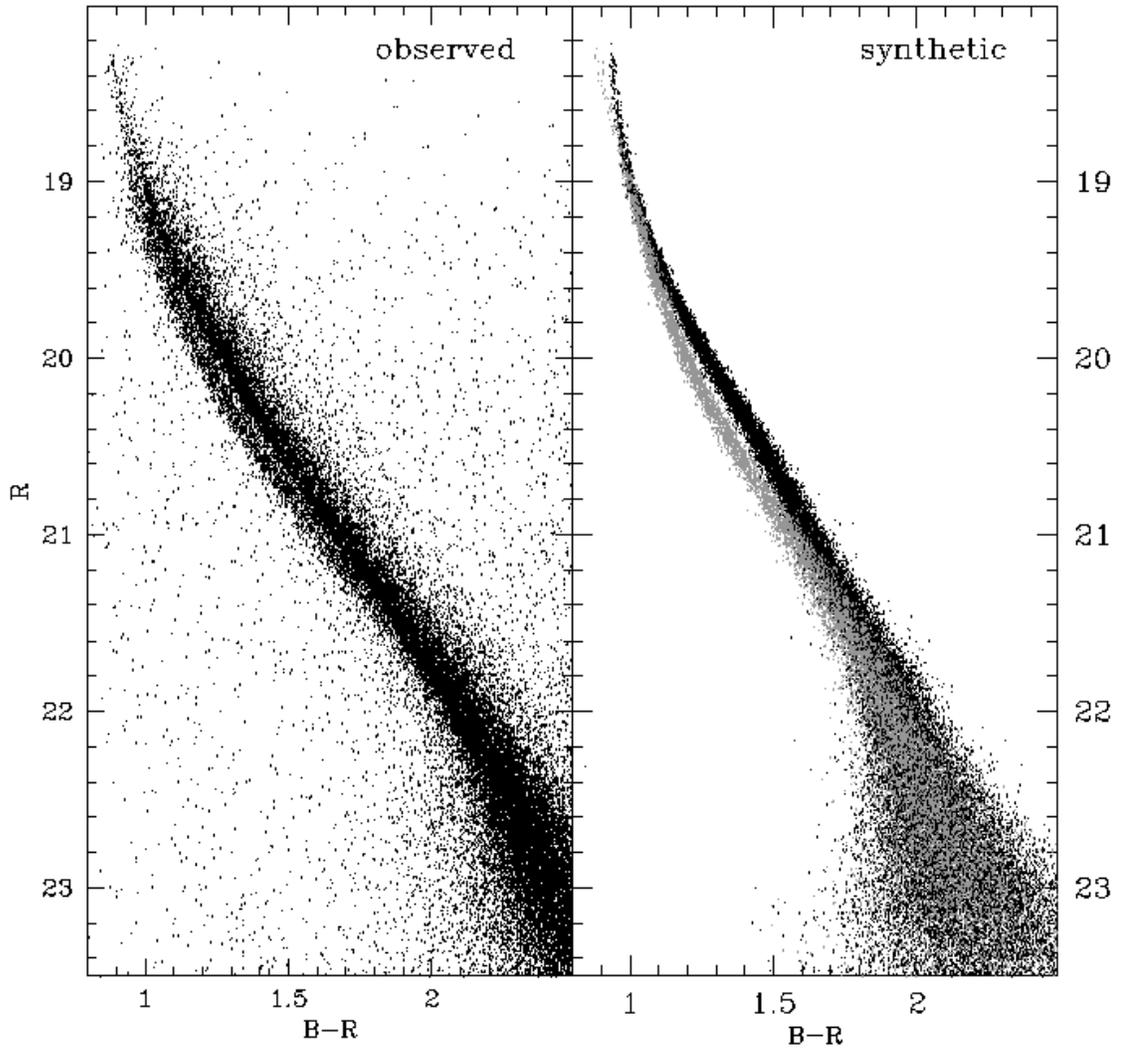}
\caption{Comparison between the observed (R, B-R) CMD of $\omega$ Cen (left $panel$) 
with a synthetic CMD (right $panel$). Black points simulate stars with [Fe/H]=-1.6
and Y=0.246, grey points simulate stars with [Fe/H]=-1.3 and  Y=0.40 .} 
\label{synth}
\end{figure}

In Fig. \ref{synth} we compare the CMD of Fig. \ref{cmdms} with a synthetic CMD
obtained adopting for the two MS components the stellar parameters listed 
above, a Salpeter (1955) Initial Mass Function and photometric errors derived 
from the arificial star technique described in Sect. 4 . Note that the synthetic CMD
reproduces quite well the observed CMD also in the faint part of the MS where the
bMS merges with the rMS at $R \sim 21$ . 
The effective temperature of the lower part of the MS is indeed less 
sensitive to a variation of the helium content with respect to the upper part 
(see Alexander et al. 1997). This occurrence can be explained with simple arguments. 
In very low mass stars ($< 0.5 M_{\odot}$),
the onset of the $H_2$ recombination (below 6000 K) induces the formation of a deep 
convective envelope characterized by a very efficient energy transport. 
In practice, the temperature gradient coincides with the adiabatic gradient up to 
the top of the convective zone. 
In this condition, since the radiative flux is 
negligible, the effective temperature is mainly controlled by the details of the 
equation of state and it is less sensitive to the variation of the radiative opacity. 
In particular, the adiabatic gradient is dominated by the presence of hydrogen 
molecules, while the atomic H and He only give minor contributions. 
Indeed, owing to the presence of these molecules, the number of internal degrees of 
freedom increases, the specific heat increases and the adiabatic gradient 
significantly decreases with respect to the typical value of a monoatomic perfect 
gas (with just 3 degrees of freedom and $dlogT/dlogP=0.4$). 
Detailed stellar models calculations confirm that the separation between isochrones 
computed with different helium abundances becomes progressively less evident at 
fainter magnitudes. The increasing
photometric error introduces a further confusion at lower magnitudes making
impossible to distinguish the two MSs. 

\section{Discussion and Conclusions}

The results presented in this paper indicate that the bMS presents peculiar
structural properties which differ from those of the dominant MS population of $\omega$
Cen. In particular, bMS stars are significantly more concentrated toward the 
cluster center. 
Note that differences in the spatial
distribution of different RGB populations were already found by Norris et al.
(1997), Pancino et al. (2003) and Sollima et al. (2005a) which showed that the
metal-rich ($[Fe/H]>-1.4$) stellar populations of $\omega$ Cen are more concentrated than the
dominant metal-poor population. Peculiarities in the radial distributions were
also found among the HB stars of $\omega$ Cen: Bailyn et al. (1992) and Rey et
al. (2004) showed that the population of Extreme Horizontal Branch (EHB) stars of $\omega$ Cen is more
concentrated toward the cluster center than MS and Sub Giant Branch (SGB) stars. 

A detailed comparison with theoretical isochrones confirms that a
significant helium overabundance ($\Delta Y > 0.10$) could explain the observed bMS
morphology, as already suggested by Norris (2004), Piotto et al. (2005) and Lee
et al. (2005).  
Note that the existence of a 
stellar population with such a large helium overabundance
can also naturally explain the HB morphology (in particular the existence of the population of EHB stars 
observed in the cluster, Lee et al. 2005), the location of the RGB bumps (Sollima et al. 2005a) and the 
SGB morphology (Sollima et al. 2005b) remaining consistent with the
short ($< 2 Gyr$) star formation time-scale estimated by Sollima et al. (2005b).
The peculiar structural properties
of the bMS indicate that the star formation process in $\omega$ Cen could 
have proceeded favoring the formation of helium-rich population(s) in the
inner region of the cluster. 

However, this scenario poses serious problems in drawing the overall chemical enrichment 
history of this stellar system. In fact, none among the known chemical enrichment
mechanisms is able to produce the huge amount of helium required
to reproduce the observed MS morphology without drastically increasing the
metal abundance. Indeed, the derived helium-to-metals abundance gradient
between the dominant metal-poor and bMS population turns out to be $\delta
Y/\delta Z >> 100$, in stark contrast with more canonical values of $\delta Y/\delta Z \sim
3-4$ (Jimenez et al. 2003). In this respect, an intriguing puzzle is presented 
by the observed incongruences between the expected and the observed luminosity of
metal-intermediate ($[Fe/H] \sim -1.2$) 
RR Lyrae found in $\omega$ Cen by Sollima et al. (2006, see also 
Rey et al. 2000 and Norris 2004). Note that a helium-rich  
population is not expected to produce a sizeable RR Lyrae component 
(Lee et al. 2005). Hence, two populations with
similar metallicities but very
different helium abundances seem to coexist within the cluster.

A selective self-enrichment process has to be invoked in order to produce the
required amount of helium while keeping the metal abundance practically unchanged.
Moreover, a very efficient mechanism is required in order to homogenize and
efficiently re-use all the ``enriched'' material ejected by the
previous generation of polluting stars.

However, a {\em pure} self-enrichment scenario may not be the best
description of the evolution of $\omega$~Cen. 
Alternative scenarios can also be considered.

In particular, the bMS 
population could have formed in a different environment, thus not partaking in the 
chemical enrichment process of $\omega$~Cen. In this 
case, a complex interplay of chemical and dynamical evolution has to be 
taken into account, including gas exchange with the Milky Way and/or minor 
mergers, within a framework such as the binary cluster mergers scenario 
(Makino et al. 1997; Minniti et al. 2004).

Bekki \& Norris (2006) noted
that the large fraction of bMS stars ($\sim 24$\%) cannot be explained without 
assuming that most of the helium 
enriched gas necessary to form the bMS originated from external sources.
They suggested that part of the helium-enriched stars formed from gas ejected by
field stellar populations surrounding $\omega$ Cen when it was the nucleus of an
ancient dwarf galaxy, and later fell into the central region of the system.

Another possible explanation can be provided by the presence of
spatial variation in the helium abundance of the protostellar clouds from which
the helium-rich population(s) of $\omega$ Cen formed.
Diffusion could produce such an effect as a result of the different acceleration
imparted to helium atoms because of their different atomic masses (Chuzhoy 2006). 

\acknowledgments

This research was supported by the Agenzia Spaziale Italiana and the Ministero dell'Istruzione, dell'Universit\`a
e della Ricerca. We warmly thank Paolo Montegriffo for assistance during the
catalogs cross-correlation and astrometric calibration process.

\clearpage

\begin{deluxetable}{llcr}
\tablewidth{0pt}
\tablecaption{Observing logs\label{obs_logms}}
\tablehead{
\colhead{Field}      & \colhead{Date}                &
\colhead{Filter}     & \colhead{Exp time}\\
\colhead{}         & \colhead{}              &
\colhead{}         & \colhead{sec}}
\startdata
1 & \rm{10~Apr~2005} & B & 1099 \\
1 & \rm{10~Apr~2005} & B & 1099 \\
1 & \rm{10~Apr~2005} & R & 394  \\
1 & \rm{10~Apr~2005} & R & 394  \\
2 & \rm{16~Mar~2005} & B & 1099 \\
2 & \rm{16~Mar~2005} & B & 1099 \\
2 & \rm{16~Mar~2005} & R & 394  \\
2 & \rm{16~Mar~2005} & R & 394  \\
3 & \rm{16~Mar~2005} & B & 1099 \\
3 & \rm{16~Mar~2005} & B & 1099 \\
3 & \rm{16~Mar~2005} & R & 394  \\
3 & \rm{16~Mar~2005} & R & 394  \\
4 & \rm{08~Apr~2005} & B & 1099 \\
4 & \rm{08~Apr~2005} & B & 1099 \\
4 & \rm{08~Apr~2005} & R & 394  \\
4 & \rm{08~Apr~2005} & R & 394  \\
5 & \rm{10~Apr~2005} & B & 1099 \\
5 & \rm{10~Apr~2005} & B & 1099 \\
5 & \rm{10~Apr~2005} & R & 394  \\
5 & \rm{10~Apr~2005} & R & 394  \\
6 & \rm{12~Apr~2005} & B & 1099 \\
6 & \rm{12~Apr~2005} & B & 1099 \\
6 & \rm{12~Apr~2005} & R & 394  \\
6 & \rm{12~Apr~2005} & R & 394  \\
7 & \rm{04~Apr~2005} & B & 1099 \\
7 & \rm{04~Apr~2005} & B & 1099 \\
7 & \rm{04~Apr~2005} & R & 394  \\
7 & \rm{04~Apr~2005} & R & 394  \\
8 & \rm{12~Apr~2005} & B & 1099 \\
8 & \rm{12~Apr~2005} & B & 1099 \\
8 & \rm{12~Apr~2005} & R & 394  \\
8 & \rm{12~Apr~2005} & R & 394  \\
\enddata
\end{deluxetable}

\clearpage

\clearpage

\clearpage

\clearpage

\clearpage

\clearpage

\clearpage

\clearpage

\clearpage

\clearpage

\end{document}